\documentclass{aa} 
\usepackage{graphicx}
\usepackage{txfonts}
\def\gapprox{{_>\atop{^\sim}}}

\def\cmmd{\rm {cm^{-3}}}
\def\cmmt{\rm {cm^{-2}}}

\def\s-1{\rm {s^{-1}}}
\def\twco{$^{12}$CO}
\def\thco{$^{13}$CO}

\def\etal {et al.}
\def\kms {\hbox{${\rm km\,s}^{-1}$}}

\def\lsun{L$_{\odot}$}
\begin{document}
\title{Overluminous HNC Line Emission in Arp~220, NGC~4418 and Mrk~231
 - Global IR Pumping or XDRs?}
\author{S.~Aalto\inst{1}, M. Spaans\inst{2}, M. C. Wiedner\inst{3}, S. H\"uttemeister\inst{4}}
\offprints{S. Aalto}
\institute{
 Onsala Rymdobservatorium, Chalmers Tekniska H\"ogskola, S - 439 92 Onsala, Sweden
 \and 
 Kapteyn Astronomical Institute, PO Box 800, 9700 AV Groningen, The Netherlands
 \and
 I. Physikalisches Institut, Universität zu K\"oln, Z\"ulpicher Str. 77, D - 50937 K\"oln, Germany
 \and
 Astronomisches Institut der Universit\"{a}t Bochum,
 Universit\"atsstra\ss{}e 150,  D - 44780 Bochum, Germany
}
\date{September 29, 2006  / November 29, 2006 }

\abstract
% context heading
{In recent studies of 3mm $J$=1--0 HNC emission from galaxies
it is found that the emission is often bright which is unexpected in warm, star forming clouds.
We propose that the main cause for
the luminous HNC line emission is the extreme radiative and kinematical environment in starburst and active nuclei.}
{To determine the underlying excitational and chemical causes behind the luminous HNC emission
in active galaxies and
to establish how HNC emission may serve to identify important properties of the nuclear source.}
{We present mm and submm JCMT, IRAM 30m and CSO observations of the $J$=3--2 line of HNC
and its isomer HCN in three luminous galaxies and $J$=4--3 HNC observations of one galaxy.
The observations are discussed in terms of physical conditions and excitation as well as
in the context of X-ray influenced chemistry.}
{The ultraluminous mergers Arp~220 and Mrk~231 and the luminous
IR galaxy NGC~4418 show the HNC $J$ 3--2 emission being brighter than the HCN 3--2 emission by
factors of 1.5 to 2. We furthermore report the detection of HNC $J$=4--3 in Mrk~231.
Overluminous HNC emission is unexpected in warm molecular gas in ultraluminous galaxies since
I(HNC)$\gapprox$I(HCN) is usually taken as a signature of cold (10 - 20 K) dark clouds.
Since the molecular gas of the studied galaxies is warm ($T_{\rm k} \gapprox 40$ K), we present
two alternative explanations to the overluminous HNC: 
a) HNC excitation is affected by pumping of the rotational levels through the mid-infrared
continuum and b) XDRs (X-ray Dominated Regions) influence the abundances of HNC.\\
HNC may become pumped at 21.5 $\mu$m brightness temperatures of $T_{\rm B} \gapprox$ 50 K,
suggesting that HNC-pumping could be common in warm, ultraluminous galaxies with compact IR-nuclei.
This means that the HNC emission is no longer dominated by collisions and its luminosity may not be
used to deduce information on gas density.
On the other hand, all three galaxies are either suspected of having buried AGN - or the presence of
AGN is clear (Mrk~231) - indicating that X-rays may affect the ISM chemistry.}
{We conclude that both the pumping and XDR alternatives imply molecular cloud ensembles distinctly
different from those of typical
starforming regions in the Galaxy, or the ISM of less extreme starburst galaxies. The HNC
molecule shows the potential of becoming an additional important tracer of extreme nuclear environments.
%We propose observational methods for distinguishing between the pumping and XDR scenario, but also find
%that both may be occuring simultaneously.} 
}
\titlerunning{Overluminous HNC in Arp~220, NGC~4418 and Mrk~231}
\maketitle

\section{Introduction}

In order to understand the AGN and star formation activity in the centres of luminous galaxies it is essential
to study the prevailing conditions of the dense
($n({\rm H}_2) \geq 10^4$\,cm$^{-3}$) molecular gas. The polar molecule HCN (dipole moment 2.98 debye)
is commonly used as a tracer of this dense gas-phase. In particular in distant
luminous (L$_{\rm IR}> 10^{11}$ L$_{\odot}$, LIRGs) and
ultraluminous (L$_{\rm IR}> 10^{12}$ L$_{\odot}$, ULIRGs) systems the HCN 1--0 line
is the prototypical tracer of 
dense gas content (e.g. Solomon \etal\ 1992, Helfer \& Blitz 1993, Curran, Aalto \& Booth 2000,
Gao \& Solomon 2004). HNC, the isomer of HCN, traces gas of equally high density. In dense, Galactic,
molecular cloud cores it has been suggested to trace gas temperature: Neutral-neutral chemical
models predict that the HCN/HNC abundance ratio increases with increasing temperature.
This is supported by the fact 
that the measured ${{\rm HCN} \over {\rm HNC}}$ abundance ratio is especially high in the vicinity of 
the hot core of Orion KL (e.g. Schilke \etal\ 1992, Hirota \etal\ 1998).

{\it It is therefore surprising that the HNC/HCN $J$=1--0 intensity ratios are found to be
low in luminous galaxies (e.g. H\"uttemeister \etal\ 1995, Aalto \etal\ 2002 (APHC02)) and that
the HNC/HCN line ratio appears to increase with galactic luminosity (APHC02).}

As an explanation to the abnormally bright HNC emission, APHC02 suggest that 
the processes are dominated by fast ion-neutral chemistry in moderately dense PDR-like regions,
instead of the neutral-neutral chemistry likely governing the hot dense cores of the Orion cloud. 
In this case, one would expect the HCN and HNC abundances and excitation to be very similar to
each other - independently of gas temperature. A study of the higher rotational transitions should
show equal brightness for the two species. 

Another possible scenario suggested by APHC02 is that, instead of being collisionally excited, HNC
is being radiatively excited.  HNC may be pumped by 21.5\,$\mu$m continuum radiation through vibrational
transitions in its degenerate bending mode. This would likely result in significant differences in
the HCN and HNC excitation. 

Alternatively, Meijerink \& Spaans (2005) suggest that the HNC abundance may become enhanced over that of HCN
in warm, dense XDRs (X-ray Dominated Regions). The deeply penetrating X-rays
induce an ionization structure that differs from the one in a PDR and that
favors asymmetries which exist in the HNC and HCN chemical pathways.
This offers a chemical explanation for the abnormal
HNC luminosity that is linked to the accreting black hole. \\

In order to investigate the underlying cause behind the bright HNC emission in luminous
galaxies, we have searched for HCN and HNC $J$=3--2 emission in a sample of LIRG and ULIRG
galaxies with the JCMT and IRAM 30m telescopes. This will allow us to investigate the excitation
of the HNC and HCN molecules. In this paper, we present the first results of
this study on the HNC 3--2 emission of two ultralumious
galaxies, Arp~220 and Mrk~231, and one luminous IR galaxy, NGC~4418.
For these galaxies we find the HNC 3--2 luminosity to be greater than that of HCN $J$=3--2 and
we discuss two possible scenarios: mid-infrared (mid-IR) pumping and X-ray dominated chemistry (XDRs).
From now on, we omit the $J$ and refer to a rotational
transition only as: 3--2 instead of: $J$=3--2.

\section{Observations and results}

We have used the JCMT telescope to measure the HNC 3--2 (271~GHz) (Figure 1 and Figure 2) lines towards
the ultraluminous galaxies Arp~220 and Mrk~231.
Observations were made in April 2005, and the system temperatures were typically 450 K.
Pointing was checked regularly on SiO masers and the rms was found to be 2$''$.
Furthermore, the HNC 4--3 line (353~GHz) was observed in Mrk~231 in February 2006.
In addition, we observed the HCN 3--2 (Figure 3) line in Mrk~231 with the CSO
telescope in 1997. With JCMT we also observed the \twco\ 2--1 line of Mrk~231 and used it
to compare with the same line observed with the CSO towards the same galaxy to compare the
intensity scales of the two telescopes. The \twco\ 2--1 line intensities from the two
telescopes were found to agree within 10\%.
The HNC and HCN 3--2 lines of NGC~4418 were observed in May and July 2006 with the IRAM 30
telescope. Pointing rms ranged between 1.´´5 and 2´´ and system temperatures ranged
between 400 and 600 K.
Beam sizes and efficiencies are shown in Table~1.  Molecular line ratios are presented in Table~2.
Note that line intensities were corrected
for beam size when compared to each other.

\begin{table}
\caption{\label{beam} Observational parameters}
\begin{tabular}{lccc}
Transition & $\nu$ [GHz] & HPBW [$''$] & $\eta_{\rm mb}$\\[4pt]
\hline
\hline \\
{\it JCMT:}\\
HCN 3--2 & 266 & 20 & 0.69\\
HNC 3--2 & 271 & 19 & 0.69\\
HNC 4--3 & 353 & 14 & 0.63\\

{\it CSO:}\\
HCN 3--2 & 267 & 30  & 0.69\\
CO 2--1 & 230 & 34 & 0.69\\

{\it IRAM:}\\
HCN 3--2 & 267 & 9.5  & 0.46\\
HNC 3--2 & 271 & 9.5  & 0.46\\
\hline \\
\end{tabular} \\
\end{table}

\begin{table} 
\caption{\label{beam} Line ratios}
\begin{tabular}{lccccc}
Galaxy & ${\rm HNC} \over {\rm HCN}$ 3--2 & HNC ${3-2 \over 1-0}$ & HCN ${3-2 \over 1-0}$\\[4pt]
\hline
\hline \\
Arp~220 & $1.9 \pm 0.3^{\rm a}$ & $1.8 \pm 0.3^{\rm b}$ & 0.9$^{\rm c}$\\
Mrk~231 & $1.5 \pm 0.2$ & $0.7 \pm 0.2^{\rm b}$ & $0.3 \pm 0.1^{\rm d}$\\
NGC~4418 & $2.3 \pm 0.3$ & $0.8 \pm 0.2^{\rm e}$ & $0.3 \pm 0.1$ \\
\hline \\
\end{tabular} \\
a) The HNC/HCN 3--2 line ratio is consistent with the one found by Cernicharo et al (2006)
b) HNC 1--0 from APCH02
c) HCN 3--2 from Wiedner \etal\ (2004).
d) HCN 1--0 data from Curran \etal\ (2000).
e) HNC 1--0 data from Monje \etal\ (in preparation). HCN 1--0 from Kohno \etal\ (2004).
The HNC 4--3/3--2 line ratio for Mrk~231 is 0.5 but the lack of baseline for the HNC 4--3 data renders this value
somewhat uncertain. Stated errors are $2\sigma$ rms errors.

\end{table}

\begin{table} 
\caption{\label{beam} Gaussian HNC and HCN line fits}
\begin{tabular}{lcccc}
Transistion &  	Arp~220 &	NGC~4418 &	Mrk~231 \\[4pt] 
\hline
\hline\\
$\int I({\rm HNC\,} 3-2)$ [K \kms] & $18.6 \pm 0.7^{\rm a}$ & $10.8 \pm 1^{\rm b}$ &
	$2.9 \pm 0.2^{\rm a}$ \\
$V_{\rm c}$ [\kms] & 5330 & 2120 & 12095\\
$\Delta V$  [\kms] & 390 & 150 & 313\\
$T_{\rm peak}$ [mK] &  44 & 63  & 9.0 \\

\hline \\
$\int I({\rm HNC\,} 4-3)$ [K \kms] & -- & -- & $2.7 \pm 0.2^{\rm a}$  \\
$V_{\rm c}$ [\kms] & -- & -- & 12144\\
$\Delta V$ [\kms] & -- & -- & 278 \\
$T_{\rm peak}$ [mK] & -- & -- &  9.0 \\

\hline \\
$\int I({\rm HCN\,} 3-2)$ [K \kms] & -- & $5.2 \pm 0.6^{\rm b}$ & $0.7 \pm 0.1^{\rm c}$ \\
$V_{\rm c}$ [\kms] 	& -- & 2120	& 12095\\
$\Delta V$ [\kms]  	& -- & 240  & 178 \\
$T_{\rm peak}$ [mK]  	& -- & 20 	& 4.2 \\
\hline \\

\end{tabular} \\
\\
a) JCMT b) IRAM 30m c) CSO \\
Integrated intensities and line intensities are in main beam brightness scale.
Note that the data are taken with different beam sizes due to the different telescopes.
This has been corrected for in the line ratio table.

\end{table}

\subsection{Arp~220}

The HNC 3--2 line is about a factor of two brighter than the HCN 3--2 line (see Table~2).
For Arp~220, we find that the HNC excitation is superthermal with a 3--2/1--0 ratio
of 1.8. Wiedner \etal\ (2004) find the HCN 3--2/1--0 ratio to be thermal with ratios
close to 0.9 (see Table~2). The HCN and HNC line widths are similar although there is 
a difference in line shape.

\begin{figure}
\resizebox{8cm}{!}{\includegraphics[angle=0]{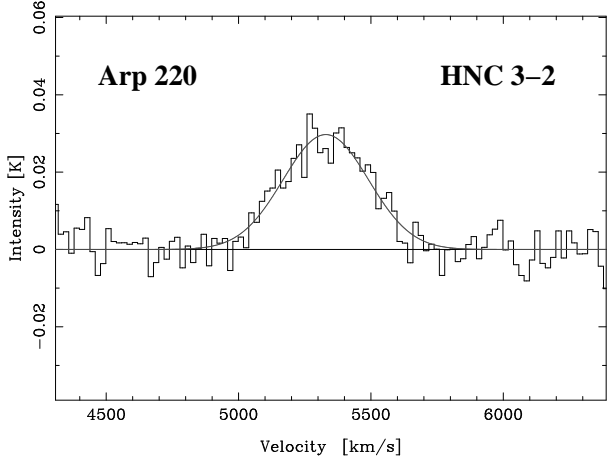}}
\caption{\label{hnc32_a220} JCMT HNC 3--2 spectrum of Arp~220. The scale is in $T_{\rm A}^*$
and should be multiplied with 1/0.69=1.45 for the scale to be transferred into main beam brightness
(see Table~1). The velocity resolution has been smoothed to 30 \kms. The Gaussian fit is presented in Table~3.}
\end{figure}

\subsection{Mrk~231}

In Mrk~231, the HNC 3--2 line is a factor of 1.5 brighter than the HCN 3--2 line (Table~2). In contrast to Arp~220
this is mainly due to the larger line width of the HNC line, while the intensities for the two lines
(corrected for the difference in beam size) is similar (Table~3).
The HNC 3--2/1--0 ratio is about 0.7 while the
HCN 3--2/1--0 ratio is 0.3 (Table~2). The HNC/HCN intensity ratio is close to
unity in the 1--0 transition (APHC02).
The HNC 4--3 line is clearly detected and the HNC 4--3/3--2 ratio is 0.5, but the lack of available baseline renders the
estimated integrated intensity somewhat uncertain.

\begin{figure}
\resizebox{8cm}{!}{\includegraphics[angle=0]{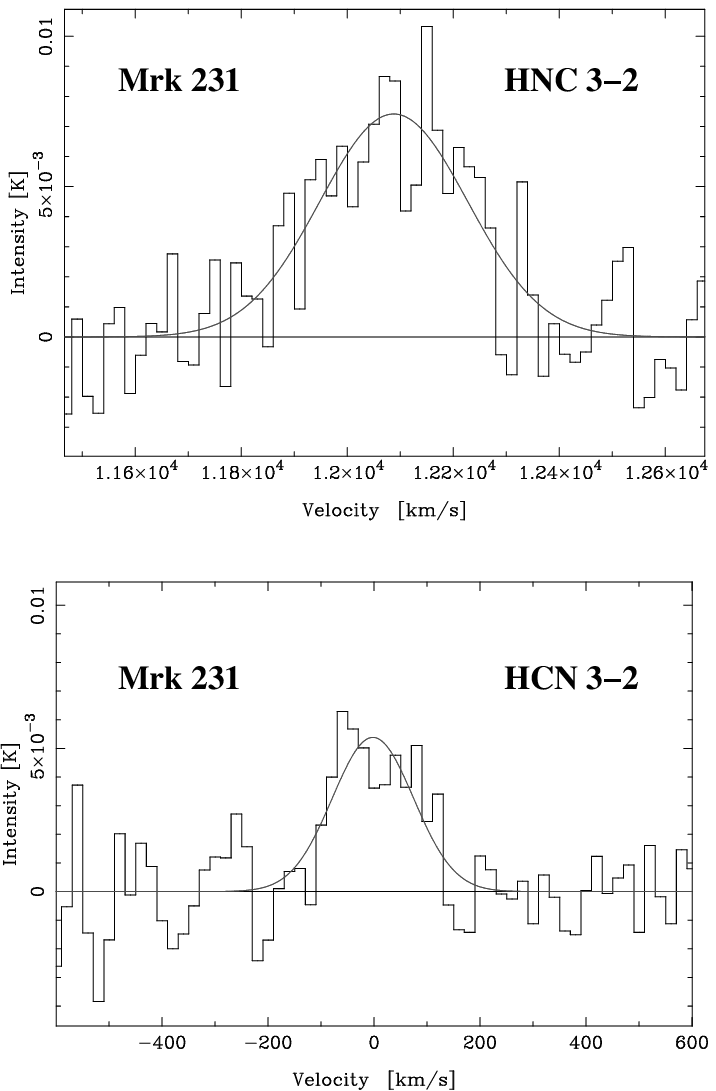}}
\caption{\label{hnc32_mrk231} Upper panel: JCMT HNC 3--2 spectrum of Mrk~231. The scale is in $T_{\rm A}^*$
and should be multiplied with 1/0.7=1.43 for the scale to be transferred into main beam brightness
(see Table~1). The velocity resolution has been smoothed to 20 \kms. Lower panel: CSO HCN 3--2 spectrum of Mrk~231. The scale is in $T_{\rm A}^*$
and should be multiplied with 1/0.69 for the scale to be transferred into main beam brightness
(see Table~1). The velocity resolution has been smoothed to 20 \kms. The velocity scales are different due to
two different observing approaches. The spectra, however, have been scaled to the same resolution and bandwidth
so that the different line widths are apparent. The total available bandwidth for the HNC spectrum is not shown
here to enable comparison with the HCN spectrum. The Gaussian fits are presented in Table~3.}
\end{figure}

\begin{figure}
\resizebox{8cm}{!}{\includegraphics[angle=0]{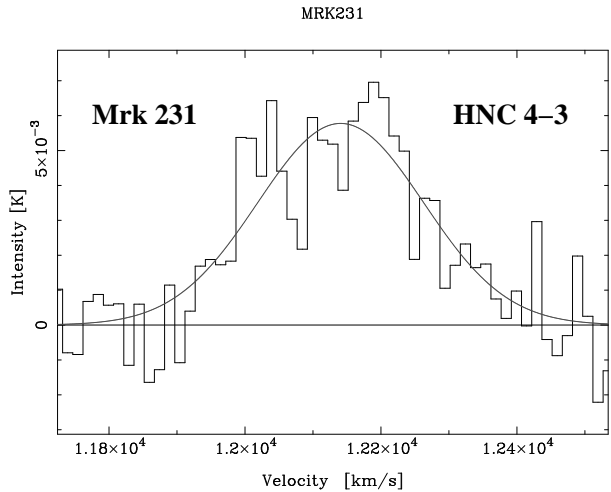}}
\caption{\label{hnc43_mrk231} JCMT HNC 4--3 spectrum of Mrk~231. The scale is in $T_{\rm A}^*$
and should be multiplied with 1/0.63=1.6 for the scale to be transferred into main beam brightness
(see Table~1). The velocity resolution has been smoothed to 20 \kms. The Gaussian fit is presented in Table~3.}
\end{figure}

\subsection{NGC~4418}

We find that the HNC 3--2 emission is brighter than the HCN emission by a factor of 2.3 (Table~2) which makes
NGC~4418 the galaxy with the largest HNC/HCN 3--2 ratio measured so far. The difference in integrated intensity
is due to the HNC line having a significantly higher intensity while its line width is narrower than
that of HCN (Table~3).
Furthermore, the excitation between HCN and HNC is distinctly different: while HCN is subthermally
excited with 3--2/1--0 line ratios of 0.3, we find that the corresponding ratio for HNC is closer to unity. 

\begin{figure}
\resizebox{7cm}{!}{\includegraphics[angle=0]{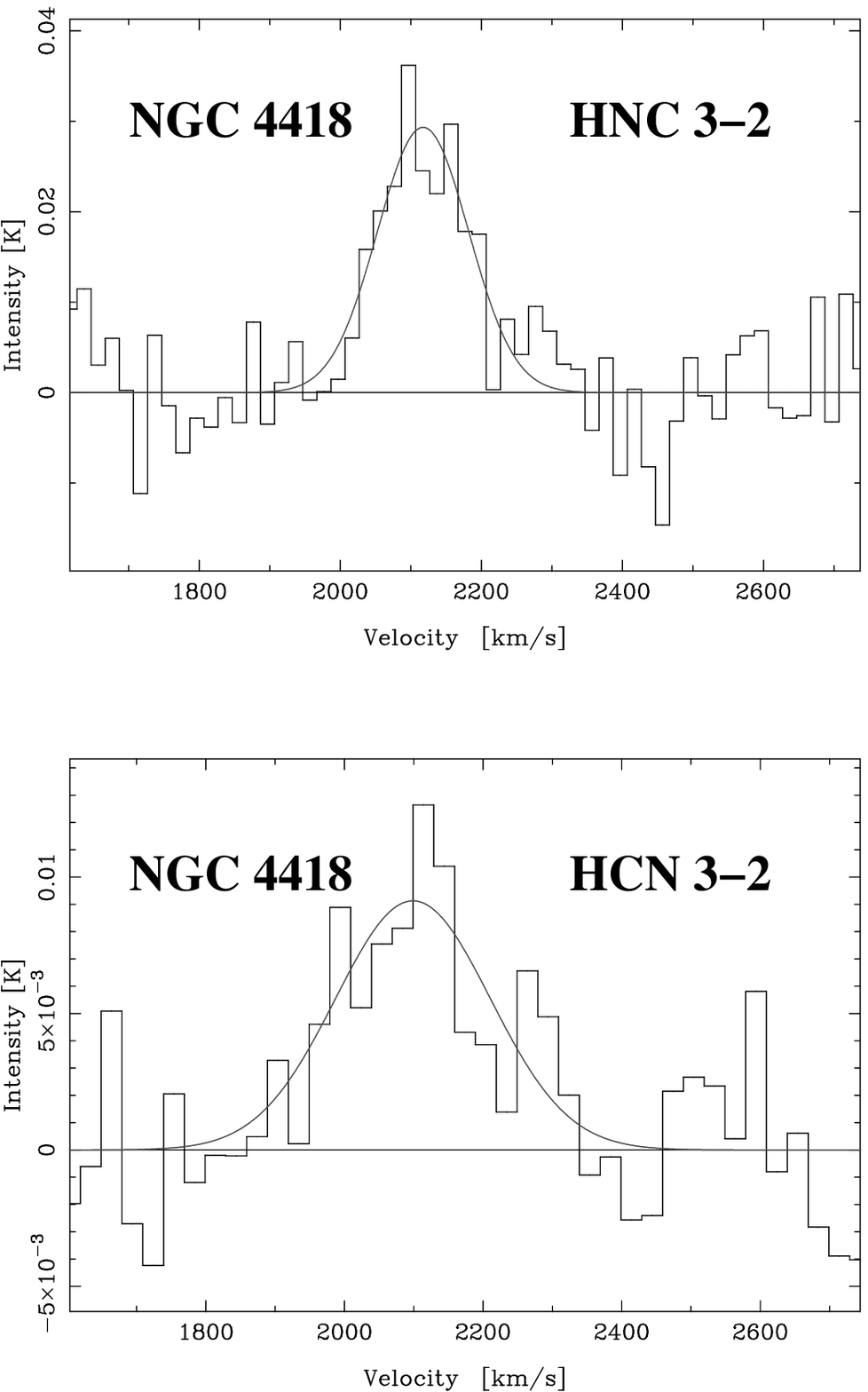}}
\caption{\label{hnc32_n4418} IRAM HNC (upper panel) and HCN (lower panel) 3--2 spectrum of NGC~4418. The scale is
in $T_{\rm A}^*$
and should be multiplied with 1/0.46=2.17 for the scale to be transferred into main beam brightness
(see Table~1.). The velocity resolution has been smoothed to 30 \kms. The Gaussian fit is presented in Table~3. }
\end{figure}

%\section{Discussion}

\begin{table} 
\caption{\label{mid-ir} Mid-IR and dust properties of Arp~220, NGC~4418 and Mrk~231}
\begin{tabular}{lcccc}
Galaxy & $L_{\rm FIR}$ & $A_{\rm V}$ & $T_{\rm K}$(gas)   & $T_{\rm D}$(dust) \\[4pt]
       & [\lsun]       & [mag]       & [K]           & [K]  \\[4pt]
\hline
\hline \\
Arp~220  & $1.1 \times 10^{12}{}^{\rm a}$ & $1000^{\rm b}$  & $\gapprox 40^{\rm b}$  & $\gapprox 85^{\rm c,d}$ \\
NGC~4418 & $8 \times 10^{10}{}^{\rm e}$ & $\gapprox 50^{\rm f}$ & - & 85$^{\rm g}$  \\
Mrk~231 & $2.3 \times 10^{12}$ & - & $>34^{\rm h}$ & 128$^{\rm i}$  \\

\end{tabular} \\

a) Aalto \etal\ (1991).
b) Sakamoto \etal\ (1999).
c) Soifer \etal\ (1999) find that the two nuclei of Arp~220 are probably optically thick at 24.5 $\mu$m.
d) Gonz\'ales-Alfonso \etal\ (2004) have modelled the far-infrared spectrum of Arp~220 as a two-component model
consisting of a nuclear region of T=106 K, effective size  of $0.''4$ and optically thick in the
far-infrared - surrounded by an extended region of size 2$''$. 
e) Roche \etal\ (1986).
f) Spoon \etal\ (2001). 
g) Evans \etal\ (2003) found that the mid-IR emitting source is compact and is consistent with a 70 pc source
of brightness temperature (dust temperature) 85 K.
h) Bryant \& Scoville (1999). 
i) Soifer \etal\ (2000) find a compact source of radius
less than 100 pc at 12.5 $\mu$m and with brightness temperature $T_{\rm B} \gapprox 141$ K.
If the IR emitting source has the same size at 25$\mu$m, the
corresponding brightness temperature is 128 K. 

\end{table}

\section{The origin of overluminous HNC}

\subsection{HNC in Galactic molecular clouds}

In ion-neutral chemical processes HCNH$^+$ will recombine
to produce either HCN or HNC with (roughly) 50\% probability. In addition, the
reaction H$_2$NC$^+$ + $e$ $\rightarrow$ HNC + H, is suggested to produce more
HNC  (e.g., Hirota \etal\ 1998).
In the Milky Way, these reactions occur mainly in dark clouds for kinetic temperatures
below 24 K. In clouds with higher temperatures, the above reactions are replaced by neutral-neutral
reactions which tend to selectively destroy HNC resulting in $X$[HNC]$<X$[HCN]. In these neutral-neutral
reactions HNC reacts with hydrogen and oxygen to form HCN, NH and CO. Thus, in Galactic, warm
star forming regions, HNC is underabundant compared to HCN (e.g. Schilke \etal\ 1992).

\subsection{HNC in luminous galaxies}

In contrast to the findings for Galactic, warm starforming regions, APHC02 found significant
HNC 1--0 luminosities in the centres of warm starburst galaxies suggesting
that high abundances of HNC were produced in warm environments. APHC02 suggested that this is
due to ion-neutral reactions persisting despite high temperatures in the gas (40-150 K). The presence
of PDRs (Photon Dominated Regions) 
could provide an environment where $X$[HNC]=$X$[HCN] and the luminosities of the two species would
be similar. 
In this paper, however, we report an {\it over}luminosity of
HNC compared to HCN by factors of 1.5 to 2.3.
{\it The PDR chemistry alone does not offer a scenario where  $I$[HNC]$\gapprox I$[HCN] and in
the following two subsections we present two mechanisms that can reproduce the observed overluminosity of
HNC: mid-IR pumping and XDR chemistry.}

\subsection{IR pumping of HNC} 

\begin{figure}
\resizebox{8cm}{!}{\includegraphics[angle=0]{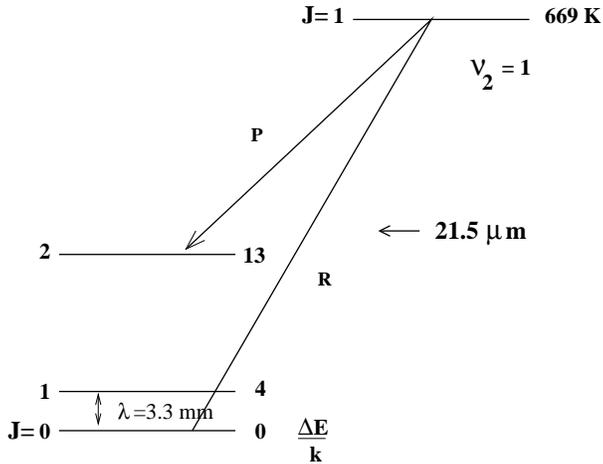}}
\caption{\label{hncex} A schematic picture of the pumping of the HNC rotational levels via the mid IR bending
transitions. The figure shows how the rotational $J$=2 level may become populated through the $\Delta J$=2 selection
rule through mid-IR pumping. The principle is the same
for higher $J$ levels.}
\end{figure}

Both HCN and HNC have degenerate bending modes in the IR. The molecule absorbs IR-photons
to the bending mode (its first vibrational state) and then it
decays back to the ground state via its $P$ branch ($\nu$=1--0, $\Delta J$=+1) or $R$-branch ($\nu$=1--0, $\Delta J$=-1) (see Figure~\ref{hncex}). In this way, a vibrational excitation may produce a change in the rotational state
in the ground level and can be treated (effectively) as a collisional excitation in the statistical
equations. Thus, IR pumping excites the molecule to the higher rotational level by a selection 
rule $\Delta J$=2.\\
For HNC, the bending
mode occurs at $\lambda$=21.5 $\mu$m (464.2 cm$^{-1}$) with an energy level $h\nu/k$=669 K and
an $A$-coefficient of $A_{\rm IR}$=5.2 s$^{-1}$. For HCN the 
mode occurs at $\lambda$=14 $\mu$m (713.5 cm$^{-1}$), energy level $h\nu/k$=1027 K and $A_{\rm IR}$=1.7 s$^{-1}$. 
{\it It is therefore significantly easier to pump
HNC, than HCN.} 
%A brief analysis shows that 
The pumping of HNC may start to become effective
when the IR background reaches an optically thick brightness temperature of $T_{\rm B} \approx$ 50 K
and the gas densities are below critical.\\

\noindent
{\it The competition between collisions and radiative excitation:}\\
The IR pumping will compete with
collisions for the excitation of HNC - when the IR field becomes more intense,
higher densities are required to successfully compete with the radiative excitation.
A simplified analysis includes comparing the rate of collisional excitation with the  
rate of mid-IR photon absorption.
A wavelength of 21.5 $\mu$m corresponds to a photon energy of $E/k$ = 669 K and
the IR pumping rate is roughly $P_{\rm IR} \propto A_{\rm IR} / (e^{669/T_{\rm B}} - 1)$ where
the $T_{\rm B}$ is the mid-IR brightness temperature (optically thick dust temperature)
and $A_{\rm IR}$ is the Einstein coefficient for the mid-IR bending transition.
For HNC, the collisional rate is roughly $2 \times 10^{-10} \, n({\rm H}_2)$ s$^{-1}$ where $n$
is in $\cmmd$.
$A_{\rm IR}$ for HNC is 5.2 s$^{-1}$, so when $T_{\rm B}$= 55 K, the pumping rate is
$2.7 \times 10^{-5}$ may dominate at
gas densities less than $10^5$ $\cmmd$. When $T_{\rm B}$ is 85 K, then
the IR pumping rate is almost 100 times faster than at 55 K. Radiative excitation
through mid-IR pumping may then dominate over collisional excitation up to a density
greater than $n=10^6$ $\cmmd$.  These rough
numbers illustrate how sensitive the competition between pumping and
collisions is to the background IR brightness. \\
For HCN, the pump rate is two orders of magnitude slower at 85 K, compared to HNC. This is
because the $A_{\rm IR}$ coefficient is lower and the energy level for the bending mode is
higher (see previous section).\\

\noindent
{\it Pumping in a two-phase molecular medium}\\
In a situation where the pumping completely dominates the excitation of HNC
the excitation temperatures of the rotational levels approach that of the
brightness temperature of the background IR field. However, in reality the
scenario is unlikely to be this simple. The molecular ISM will, for example, consist of a
range of densities - also in the nucleus of the galaxy.\\
Consider a simple molecular ISM where the dense ($n\gapprox 10^4$ $\cmmd$) cores in the galaxy nuclei are
surrounded by lower density ($n= 10^2 - 10^3$ $\cmmd$), unbound gas (a ``raisin roll´´
scenario - e.g. Aalto \etal\ 1995, Downes and Solomon 1998,
H\"uttemeister and Aalto 2001, Aalto 2005). In this scenario, the diffuse gas contributes a significant
fraction of the lower-$J$ \twco\ emission and has a high volume filling factor.
In the case of an intense mid-IR field, all of the
HNC molecules exposed to it will be dominated by the radiative excitation. In a more dilute
field, with $T_{\rm B}=50$ K, for example, the dense cores may be unaffected while the HNC molecules
in the more diffuse gas may be experiencing radiative excitation. {\it Thus, the global
HNC luminosity may be significantly affected
by pumping, even if the mid-IR field is moderate}. In this scenario one would expect a radial
gradient where the effect of the radiative pumping affects a larger fraction of the gas closer to the
nucleus of the galaxy.

The presence of such a diffuse, lower density
($n= 10^2 - 10^3$ $\cmmd$) molecular medium is suggested to be present in Arp~220:
A large \twco/\thco\ 1--0 line intensity ratio ($>$30 Aalto \etal\ 1991) implies a low to moderate
optical depth of the \twco\ line. Together with subthermally excited low-$J$ CO lines, this
suggests the presence of a large filling factor of diffuse molecular gas (Aalto \etal\ 1995).
The concept of diffuse gas in ultraluminous galaxies - among them Arp~220 
- was extensively modelled and discussed by Downes and Solomon (1998).\\
For Mrk~231, a similar diffuse gas phase may be suggested by
the large CO/\thco\ 2--1 line intensity ratio (Glenn \& Hunter 2001) although further observations
and radiative transfer modeling is necessary. For NGC~4418, the presence of diffuse molecular gas
is unexplored and multi-wavelength CO and \thco\ modeling is required to address this issue.

\subsection{XDR chemistry}

The X-ray irradiation of molecular gas leads to a so-called
X-ray dominated region (e.g., Maloney \etal\ 1996, Lepp \& Dalgarno 1996)
similar to PDRs associated with bright UV sources
(Tielens \& Hollenbach, 1985). The more energetic (1-100 keV)
X-ray photons penetrate large columns ($10^{22}-10^{24}$ $\cmmt$) of gas and
lead to a different ion-molecule chemistry. Species like C, C$^+$ and CO
co-exist (unlike the stratified PDR) and the abundances of H$^+$ and He$^+$
are much larger. Molecules like H$_2$O and OH can be formed more efficiently
because molecular gas resides at higher temperatures, a consequence of the
fact that ionization heating (rather than photo-electric heating in PDRs)
dominates and is more efficient (about 70\%).
Models of XDRs by Meijerink \& Spaans (2005) indicate that
the HNC/HCN column density ratio is elevated, and reaches a value of $\sim 2$,
for gas densities around $10^5$ $\cmmd$.
This while PDRs and quiescent cloud regions exhibit ratios of unity
or less. Hence, ULIRGS that contain an AGN and possess high gas densities
are likely candidates for overluminous HNC emission (see Meijerink,
Spaans \& Israel 2007 for details).

\begin{figure*}
\resizebox{16cm}{!}{\includegraphics[angle=0]{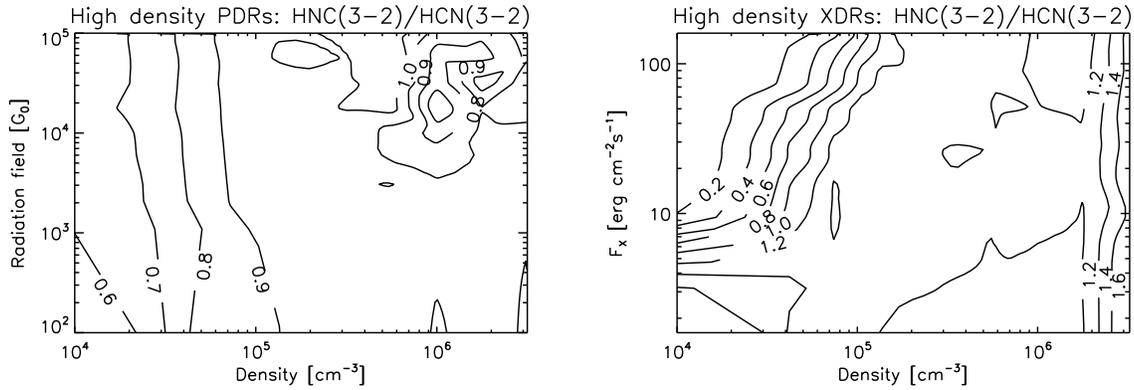}}
\caption{\label{hnc32_a220} Model line ratio HNC/HCN 3--2: PDRs (left panel) and
XDRs (right panel). Note that PDR models cannot reproduce ratios larger than unity,
whereas XDRs can. All models are for a 1 pc cloud.}
\end{figure*}

In order to compare with observations Figure 6 presents a grid of one dimensional
single-sided PDR/XDR slab
models in density and FUV/X-ray irradiation for HNC/HCN 3--2 line ratios
(Meijerink, Spaans and Israel 2007).
The model clouds have a size of 1 pc, but we have verified that our choice of
cloud size does not influence the computed line intensity ratios. We have
assumed that an individual cloud in a galaxy has a velocity line width of 5
km/s. The
impinging X-ray flux follows a $-0.9$ powerlaw in energy between 1 and 100
keV, typical of a Seyfert nucleus.
Recall here that the total flux in the Habing field (between 6 and
13.6 eV) is $1.6\times 10^{-3}$ erg s$^{-1}$ cm$^{-2}$. Hence, compared to
a PDR, we consider enhancement factors of the ambient radiation field of
$10^3-10^5$, typical of gas exposed to an AGN (Maloney \etal\ 1996,
Meijerink \& Spaans 2005).
HCN and HNC are formed in equal amounts through
dissociative recombination of HCNH$^+$ (branching ratio 50-50).
In XDRs, the degree of ionization is much larger than in PDRs.
Hence, reactions with, e.g., He$^+$, H$^+$, C$^+$ and many others, that
lead to the formation and destruction of HCN and HNC in an asymmetric way
(with HCNH$^+$ and H$_2$NC$^+$ as intermediates), are much more important.
HNC is particularly favored over HCN for moderate, $1-10$ erg s$^{-1}$ cm$^{-2}$, values of the
impinging flux, at an ambient density of
$n\ge 10^5$ cm$^{-3}$, where the abundance of ions exceeds that in a PDR
by about an order of magnitude (Meijerink, Spaans \& Israel 2006, their Figure 1).
Note that the different abundance gradients that occur for HCN and HNC
(Meijerink \& Spaans 2005, their Figure 10) at least provides a
necessary condition for the HCN and HNC line intensity profiles to differ,
i.e., to sample different kinematic regions in the nucleus of an active galaxy.

In PDRs, the ionization degree is modest and set by the ambient cosmic
ray ionization rate. Consequently, the HNC/HCN emission line ratios level off
to unity at large, $>10^{22}$ cm$^{-2}$, depths into the cloud.
Note that a strongly elevated cosmic ray ionization rate is expected in systems
that exhibit a large
star formation rate, like the 250 M$_\odot$ yr$^{-1}$ for Arp~220, because the
supernova rate will scale up
proportionally for a Salpeter IMF.
The models of Meijerink \etal\ (2006) show that even a cosmic
ray rate of $5\times 10^{-15}$ s$^{-1}$ (roughly 200 M$_\odot$ yr$^{-1}$) does
not allow cosmic ray boosted PDRs to behave as genuine XDRs in terms of
the HNC/HCN ratio.
This is basically because the very high energies of cosmic rays render their
radiative interactions (cross sections scale as $1/E^3$) with atoms and molecules
rather weak.

A top-heavy (non-Salpeter) IMF would lead
to a higher ionization rate for a fiducial star formation rate of 200 M$_\odot$ yr$^{-1}$.
Still, even for ionization rates as large as $3\times 10^{-14}$ s$^{-1}$, which would accommodate
many extreme IMFs, cosmic-ray enhanced PDRs would still not mimic XDRs in terms
of HNC, HCN and HCO$^+$. In particular, HNC/HCN would still be limited to unity
for the very high cosmic ray ionization rate mentioned above.
The reason is that X-rays couple much better to the gas than cosmic rays, i.e.
their energies are lower, while large columns of gas can still be penetrated
by them (unlike by UV photons), leading to an active ion-molecule chemistry
that favours HNC and HCO$^+$.

Finally, we note that the maximum HNC/HCN 3--2 line ratio of 1.6 in our model
clouds is smaller than the abundance ratio of about 2 found by Meijerink
\& Spaans (2005). This is a consequence of optical depth effects in the lines,
i.e., one typically reaches a $\tau =1$ surface before the full HNC/HCN abundance
gradient is sampled. A larger line width than 5 km/s for individual model
clouds would alleviate this.

\section{Discussion}

With existing data it will  be difficult to distinguish between the two
proposed scenarios of IR-pumping and XDRs. The reason for this is that the
densities required for HNC abundance enhancements in an XDR are high, $n > 10^5$ $\cmmd$
(see previous section), resulting in a
high excitation of the molecule.  This is also the case for the IR-pumping, since
the radiative excitation helps populating higher $J$-levels.

\subsection{A pumping scenario: Observational confirmation} 

The mid-IR and dust properties of the three observed galaxies are
summarized in Table~4 and
the requirements for pumping of HNC, discussed in 3.3, appear
fulfilled in the nuclei of the
galaxies. Below is a list of
useful observational tests for IR pumping:

\begin{itemize}
\item{{\it Line ratios and excitation}} If the HNC/HCN line intensity ratio exceeds the theoretical 
abundance limit of 2 from an XDR then it is necessary to invoke
radiative pumping to
explain the observed line intensity ratio. This is for instance the case
for NGC~4418 (see section 4.3.2.) 
Furthermore, a significant difference in excitation between HCN and HNC would 
be a strong indication of radiative excitation of HNC, for instance if the
higher transition
HNC lines are more luminous than those of HCN. Both molecules have similar
critical densities,
so a more excited HNC may well indicate radiative pumping. 

\item{{\it Vibrational absorption and emission lines:}}
Searching for the 21.5 $\mu$m absorption line that occurs when the HNC molecule 
absorbs the mid-IR continuum is a potential useful test. 
A possible caveat, however, is that the absorption lines may become
filled-in by emission at similar wavelengths when the molecule falls back into the
$v$=0 state.
It is also possible to search for rotational transitions of vibrationally excited HNC
in the millimeter/submillimeter range.
They are, however, often close enough in frequency
to the rotational bands that they become blended with them (they are also weaker than the purely
rotational bands) therefore sensitive, high spectral resolution observations are required. Vibrationally excited
HNC has been detected in the Galaxy towards the proto-planetary nebula CRL 618.
The lines are emerging in the hot PDR between the inner H~II region and the molecular envelope
(Schilke et al. 2002).

\item{{\it High resolution studies:}} 
High resolution, interferometric studies of HCN and HNC nuclear line emission
will show the relative distribution and brightness of the two species. 
Mid-IR observations will reveal how the infrared continuum is distributed compared to
the molecular line emission. 
Multi-transition observations of other high density tracer molecules at high resolution 
will provide an independent density measurement of the nuclear gas. One must select molecules
that are unlikely to be affected by radiative excitation (such as CS).
\end{itemize}

\subsection{An XDR scenario: Observational confirmation}

An HNC/HCN ratio larger than unity may instead be an abundance effect caused by the presence
of an XDR - rather than a PDR combined with radiative excitation (Meijerink \& Spaans 2005,
Meijerink, Spaans \& Israel 2006). The ambient
densities, $10^5$ cm$^{-3}$, and X-ray fluxes, $1-10$ erg s$^{-1}$ cm$^{-2}$,
that are required for the XDR interpretation of the Arp~220, NGC~4418 and
Mrk~231 data are quite reasonable, and only require the presence of an AGN
and an absorbing hydrogen column of $N_H>10^{22.5}$ cm$^{-2}$.
However, the alternative of IR pumping cannot be excluded on these grounds.
Below follows a list of observational tests for an XDR:

\begin{itemize}

\item{{\it HCN/HCO$^+$:}}
One can look for modest, $<1$, HCN/HCO$^+$
1--0, 2--1, etc.\ line ratios. The HCO$^+$ abundance is expected to increase
more in XDRs than the HCN abundance and this leads to HCN/HCO$^+$ rotational
emission line ratios that are less than unity, as follows.
A high X-ray flux leads to more destruction of HCO$^+$ at
the edge of an XDR, through dissociative recombination, but the large
degree of ionization, and the associated ion-molecule chemistry, more than
compensates for this at hydrogen columns in excess of $10^{23}$ cm$^{-2}$
and/or flux-to-density ratios of less than $10^{-3}$ erg s$^{-1}$ cm
(Meijerink \& Spaans 2005). That is, penetrating X-rays, much
like cosmic rays, maintain a large abundance of H$_3^+$. This ion reacts
with O to form OH$^+$ and leads to H$_3$O$^+$ through reactions with H$_2$.
Dissociative recombination then leads to OH (and water) which can react
with C$^+$ to form CO$^+$ (and HCO$^+$), subsequent reaction with H$_2$
leads to HCO$^+$. Additional
pathways exist where OH, and H$_2$O, form directly through endoergic
neutral-neutral reactions O + H$_2$ and OH + H$_2$ followed by reaction
with C$^+$. Also, once CO is abundant, reaction with H$_3^+$ leads to
HCO$^+$ directly. Note in this that the often cited Lepp \& Dalgarno
(1996) XDR work clearly shows that HCO$^+$ enjoys large abundances over
a wider range of ionization rates than HCN does, favoring HCO$^+$
over HCN when a strong source of ionization plays a role.
One may wonder about the influence of cosmic rays, as a source
of ionization, on the structure of PDRs.\\
Model 3 of Jansen \etal\ (1995) shows that raising the cosmic ray
ionization rate by a factor of 4 over the Galactic value, and thus
boosting the level of ions, enhances the HCO$^+$ abundance
by more than a factor of three, while HCN and HNC benefit only on the
factor of two level. Such a high cosmic ray rate PDR behaves somewhat
like an XDR, but is less extreme (HCN/HCO$^+$ $\sim 1$ or a bit less),
as discussed in Meijerink \etal\ (2006). The latter authors also show
that a regular PDR with a low cosmic ray ionization rate, which would
apply when no supernovae have gone off yet, leads to HCN/HCO$^+ >$ 1 for
densities of $\sim 10^5$ cm$^{-3}$ and up.\\
Thus, one can make a clear comparison between low cosmic ray rate PDRs
and XDRs {\it if} the PDR component does not occupy a larger area in the
beam. The XDR/AGN contribution loses to the star formation/FUV signal if
the latter is more spread out and care should therefore be taken with the
interpretation of correlations between nuclear activity
and elevated HCN/HCO$^+$ ratios.\\
Gracia-Carpio \etal\ (2006) find elevated
HCN/HCO$^+$ 1--0 line ratios in ULIRGs - and in general they find that the line ratio
correlates with FIR luminosity.
They attribute this to the presence of an AGN where the X-rays affect the chemistry
to impact the HCN/HCO$^+$ abundance ratio. 
However, as we discuss in this paper, for dense XDRs ($n > 10^5$ $\cmmd$) one would not expect
the HCN to be more abundant than HCO$^+$. It is possible that the findings of Gracia-Carpio
\etal\ can be explained by excitation effects (e.g. IR-pumping) or by other chemical
processes affecting abundances.

\item{{\it Very high $J$ CO emission:}}
Unambiguous diagnostics are provided by highly excited ($J>10$) rotational
lines of CO (Meijerink \etal\ 2006). XDRs more easily produce very warm CO
compared to PDRs (also PDRs with high cosmic ray rates). This is because CO is
present already at the warm XDR edge, initiated through C$^+$ + OH route above,
while the higher UV flux at the edge of
PDRs, for the same density and total energy input, suppresses CO.
These high $J$ CO lines will be accessible to the HIFI instrument on Herschel
- and to instruments on SOFIA - and can already be observed in absorption in
the near-infrared with Subaru. A combination of overluminous HNC emission and very high $J$
CO emission (or absorption) would be compelling evidence for an XDR/AGN.

\end{itemize}

\subsection{Pumping or XDRs in Arp~220, NGC~4418, Mrk~231}

{\bf Arp~220} is an ultraluminous gas-rich merger of two galaxies where the
two nuclei are separated by $\sim$300 pc (Graham \etal\ 1990). The nature
of the nuclear power source of Arp~220 is yet not established (see for instance
Sakamoto \etal\ (1999) section 7 for a discussion) since the activity is deeply
enshrouded by dust. Regardless of whether the intense infrared radiation is
driven by the nuclear starburst alone - or whether it is boosted by an AGN - the
mid-IR properties of Arp~220 fulfill the
criteria for 21.5 $\mu$m HNC pumping (see Table~4). Sakamoto \etal\ (1999) find that
a substantial fraction of the molecular gas can be found in compact, rotating nuclear structures,
which should be closely associated with the mid-infrared sources.
Even though intense emission from molecules such as CS, CN, HCO$^+$ 1--0, HCN
and HC$_3$N (Radford \etal\ 1990, Solomon \etal\ 1992), Wiedner \etal\ 2004, APCH02 indicate
gas densities in excess of $10^4$ $\cmmd$, the brightness temperature of the mid-IR field
is large enough to compete with collisions at densities up to at least $10^5$ $\cmmd$.
The high excitation of HNC, as compared to HCN, is an expected
pumping-signature since it should result in elevated excitation temperatures (also of the
higher transitions) compared to those expected
from typical collisional excitation.\\

{\it For Arp~220, the observed HNC/HCN 3--2 line ratio of 1.9 is close to the theoretical maximum of 2 for
the abundance ratio in an XDR (section 3.4) and all of the HNC and HCN 3--2 emission must emerge from
an XDR with optically thin emission for an XDR to suffice in explaining the line ratio. 
Thus, we propose that the HNC emission is affected by pumping in Arp~220 with the possible 
additional effect of an XDR. This would require Arp~220 to host a buried AGN, but such an AGN
component does not have
to be energetically dominant.}\\

{\it Pumping of HCN?}: It is also possible that part of the HCN emission
is affected by radiative excitation as well, but to a lesser degree. The requirements for the
14 $\mu$m field to be intense enough to
compete with collisions for the HCN molecule are greater for HCN than the requirements to pump
HNC - and are fulfilled in a smaller region of the western
nucleus of Arp~220. 
Barvainis \etal\ (1997) suggested that the excitation of the
HCN emission of the Cloverleaf Quasar may be affected by absorption of 14 $\mu$m continuum.
Clearly, the excitation of HCN in Arp~220 warrants further study.\\

Even though {\bf NGC~4418} is far from being an ultraluminous galaxy, several of its mid-IR and molecular
properties appear remarkably similar to those of Arp~220 (Table~4). Both galaxies exhibit rich
molecular chemistry similar to that of a Galactic hot core (e.g., APCH02, Monje \& Aalto in preparation,
Martin-Pintado et al. in preparation).
The far-infrared surface brightness is estimated to be extremely high (Evans \etal\ 2003)
and most of its molecular gas is concentrated in the inner kpc (Dale \etal\ 2005). It is unclear what
is driving the IR luminosity
of NGC~4418. Imanishi \etal\ (2004) report that the estimated star formation luminosity from the
observed PAH emission can account for only a small fraction of the infrared luminosity thus suggesting
that NGC~4418 may be largely AGN-driven. They furthermore report an elevated HCN/HCO$^+$ 1--0 line
intensity ratio which they suggest is due to the presence of an AGN-powered XDR-chemistry, although
depth-dependent models by Meijerink \& Spaans (2005) and Meijerink \etal\ (2006) seem to suggest an
opposite trend, with HCN/HCO$^+$ 1--0 $<1$ under X-ray irradiation (see section 4.2.1).
Properties presented in Table~4 show that conditions for nuclear IR pumping of HNC are fulfilled.
The HNC 3--2/1--0 line ratio is not globally superthermal - as is the case for Arp~220 -
but HNC is significantly more highly excited than HCN.\\

{\it For NGC~4418, the observed HNC/HCN 3--2 line ratio of 2.3 exceeds the theoretical maximum of 2 for
the abundance ratio in an XDR (section 3.4) and we conclude that mid-IR pumping is necessary
to explain the observed HNC/HCN 3--2 line ratio. However, an additional effect from an XDR is
likely and should be investigated with the proposed observational methods. }\\

{\bf Mrk~231} is an ultraluminous Seyfert~1 galaxy with a powerful starburst surrounding the AGN.
Davies \etal\ (2005) have studied the starburst and found a warped starburst disk
of dimensions 200 pc - with a high stellar velocity dispersion of 100 \kms\ (recent
results indicate that velocity dispersions of ULIRGs may in general be even higher (Dasyra \etal\
(2006)). 
Vibrationally excited
H$_2$ lines are found throughout the starburst disk and the overall ISM properties of the disk
must be highly unusual. The presence of an AGN driving the Seyfert~1 characteristics is well
established in Mrk~231 - in contrast to the other two galaxies in the sample.
However, Mrk~231 is not as clear a case for HNC overluminosity as Arp~220 or NGC~4418 - even though
nuclear conditions should be favourable for pumping as indicated in Table~4. The HNC 3--2 luminosity is
a factor of 1.5 greater than the HCN 3--2 luminosity - but this is largely due to a difference in
line width, not peak brightness as is the case for Arp~220 and NGC~4418. The HCN line width agrees
with that of \twco\ 2--1  and is consistent with the
rotational velocities expected from the near-face-on 1 kpc disk. The HNC 3--2 (and HNC 4--3) line width
is, however, almost a factor of two greater than that of HCN and \twco. 
The HNC 3--2 line peaks at the same velocity as the OH megamaser emission (e.g. Richards \etal\ 2005)
and has a similar line width. This may imply that the HNC 3--2 emission is more nuclear than
the HCN (although we require high resolution information to confirm this). 
In this scenario there will be a radial dependence in the intensity ratio, where the HNC/HCN line intensity
ratio may be close to unity in the disk, and then increases towards the nucleus.\\
HCN is subthermally excited, suggesting overall densities between $10^4$ and $10^5$ $\cmmd$ and
the gas kinetic temperatures are high as suggested by the interferometric observations (see Table~4).
HNC is more highly excited, suggesting it is emerging from higher density or warmer gas - or that
it is affected by radiative excitation. Both these suggestions are consistent with HNC being more
centrally concentrated than HCN. \\

{\it The observed HNC/HCN 3--2 line ratio of 1.5 is less than the theoretical maximum of 2 (section 3.4)
for an XDR and with existing data we cannot distinguish between pumping and X-ray chemistry as the main
source behind the overluminosity of HNC.}

\subsection{The extreme nuclear ISM}

Regardless of whether it is mid-IR pumping or X-ray induced chemistry (or
a combination of both) that is behind the overluminous HNC 3--2 emission, it
is clear that the prevailing {\it ISM conditions are unusual compared to similar
scales in more normal galaxies. The buried nuclear activity affects the properties
of the ISM on scales
large enough to create global effects on observed molecular line ratios.}

When comparing the HNC 1--0 luminosities of more nearby galaxies studied
by H\"uttemeister \etal\ (1995) APCH02 find that the relative HNC 1--0 luminosity
appeared to increase with luminosity - even if this effect
is small. In general, normal nearby galaxies have HNC/HCN 1--0 line ratios well
below unity. Both APCH2 and H\"uttemeister \etal\ find that HNC/HCN 1--0 line ratios
of unity (or exceeding unity) are rare.

Both Arp~220 and NGC~4418 show deep silicate absorption in their mid-IR spectra
(e.g., Spoon \etal\ 2001) suggesting visual extinction exceeding 50 magnitudes. 
It is interesting to note that it is these two galaxies that show the greatest
overluminosity in HNC - and not the Seyfert~1 galaxy Mrk~231. It is tempting to suggest
that a deeply buried, compact and highly intense source heats the large masses
of dust surrounding the nuclei of Arp~220 and NGC~4418 and that the re-radiated
IR emission is pumping the HNC emission. 
It is quite possible that we have nuclear XDRs in NGC~4418 and in Arp~220, but we note
that their effects are likely boosted by mid-IR pumping since the observed 
HNC/HCN 3--2 line ratios exceed the theoretical maximum of 2 for the HNC/HCN abundance ratio
from an XDR.  \\
Future, high resolution mm observations, and further studies in
the mid-IR and high energy X-rays will help to reveal which one of the two suggested
mechanisms is the dominant one behind the overluminous HNC 3--2 emission.

\section{Conclusions}
\begin{enumerate}

\item We have detected surprisingly bright HNC 3--2 emission in
the ultraluminous galaxies Arp~220 and Mrk~231 and in the IR luminous galaxy
NGC~4418. In all three cases
the HNC 3--2 emission outshines the HCN 3--2 emission by factors 1.5 - 2.3.
We also detected HNC 4--3 emission in Mrk~231 which shows that HNC line emission continues
to be bright also at higher transitions.

\item We propose that the HNC emission is pumped by 21.5 $\mu$m IR
continuum emission via its degenerate bending mode. This means that the
emission is no longer dominated by collisions and its luminosity may not be
used to deduce information on gas density. Ways to test for pumping include
high resolution studies, searches for the mid-IR absorptions lines and the
rotational-vibrational HNC lines.

\item Alternatively, X-rays produced by an embedded AGN can influence the
chemistry in favor of HNC, with HNC/HCN emission line ratios of 1-2. This
may produce an overluminosity of HNC without the effects of pumping.
Small HCN/HCO$^+$ line intensity ratios and, in particular, the detection
of very highly excited CO ($J\ge 10$) would confirm the XDR interpretation
of the overluminous HNC emission in Arp~220, Mrk~231 and NGC~4418. For
NGC~4418 however, the observed HNC/HCN 3--2 line ratio exceeds the theoretical
maximum for XDR chemistry, and it is necessary to invoke mid-IR pumping to 
explain the observed line ratios.

\item We conclude that for all three galaxies the molecular ISM is in an extreme
state - where the the physical conditions and/or chemistry of the gas is heavily influenced by
the radiation field from the central activity. The ISM properties are very different
from those of typical Galactic starforming molecular clouds.

\end{enumerate}

\acknowledgements{Many thanks to the JCMT and IRAM staff for their support during observations. 
Molecular databases
that have been helpful include The Cologne Database and the molecular database of Crovoisier.
We are grateful for discussions with J. Black, A.E.H. Olofsson and R. Meijerink, and for help
with observations by E. Olsson, R. Monje and K. Torstensson. Our most heartfelt thanks to D. Lis
of Caltech who helped us find and extract the Mrk~231 HCN 3--2 data from the CSO.}

\end{document}